\RequirePackage{xcolor}
\documentclass[journal,twoside,web]{ieeecolor}
\usepackage{jsen}
\usepackage{cite}
\usepackage{amsmath,amssymb,amsfonts}
\usepackage{algorithmic}
\usepackage{textcomp}
\usepackage{wrapfig}
\usepackage{tcolorbox}
\usepackage{booktabs}
\usepackage{multirow}
\usepackage{longtable}
\usepackage{makecell,array}
\usepackage{rotating}
\usepackage{adjustbox}
\usepackage{moresize}
\usepackage{soul}

\def\BibTeX{{\rm B\kern-.05em{\sc i\kern-.025em b}\kern-.08em
    T\kern-.1667em\lower.7ex\hbox{E}\kern-.125emX}}
\definecolor{abstractbg}{rgb}{0.89804,0.94510,0.83137}
\begin{document}
\title{Sparse Wearable Sonomyography Sensor-based Proprioceptive Proportional Control Across Multiple Gestures}
\author{Anne Tryphosa Kamatham, \IEEEmembership{Student Member, IEEE}, Kavita Sharma, and Srikumar Venkataraman, Biswarup Mukherjee,\IEEEmembership{Senior Member, IEEE}
\thanks{This manuscript was compiled on \today. This work has been funded in part by the Science and Engineering Research Board (SERB), Department of Science and Technology, Government of India, through a Core Research Grant (CRG/2021/004967, PI: BM), IEEE Instrumentation and Measurement Society Graduate Student Fellowship 2022 (Recipient: ATK) and by IIT Delhi through a Faculty Interdisciplinary Grant (MI02374).}
\thanks{Anne Tryphosa Kamatham and Kavita Sharma are with the Centre for Biomedical Engineering, Indian Institute of Technology Delhi, New Delhi 110016.}
\thanks{Srikumar Venkataraman is with the Department of Physical Medicine and Rehabilitation, All India Institute of Medical Sciences, New Delhi 110029.}
\thanks{Biswarup Mukherjee is with the Centre for Biomedical Engineering, Indian Institute of Technology Delhi, New Delhi 110016 and with the Department of Biomedical Engineering, All India Institute of Medical Sciences, New Delhi 110029.}
}
\IEEEtitleabstractindextext{%
\fcolorbox{abstractbg}{abstractbg}{%
\begin{minipage}{\textwidth}%
\begin{wrapfigure}[12]{r}{3in}%
\includegraphics[width=3in]{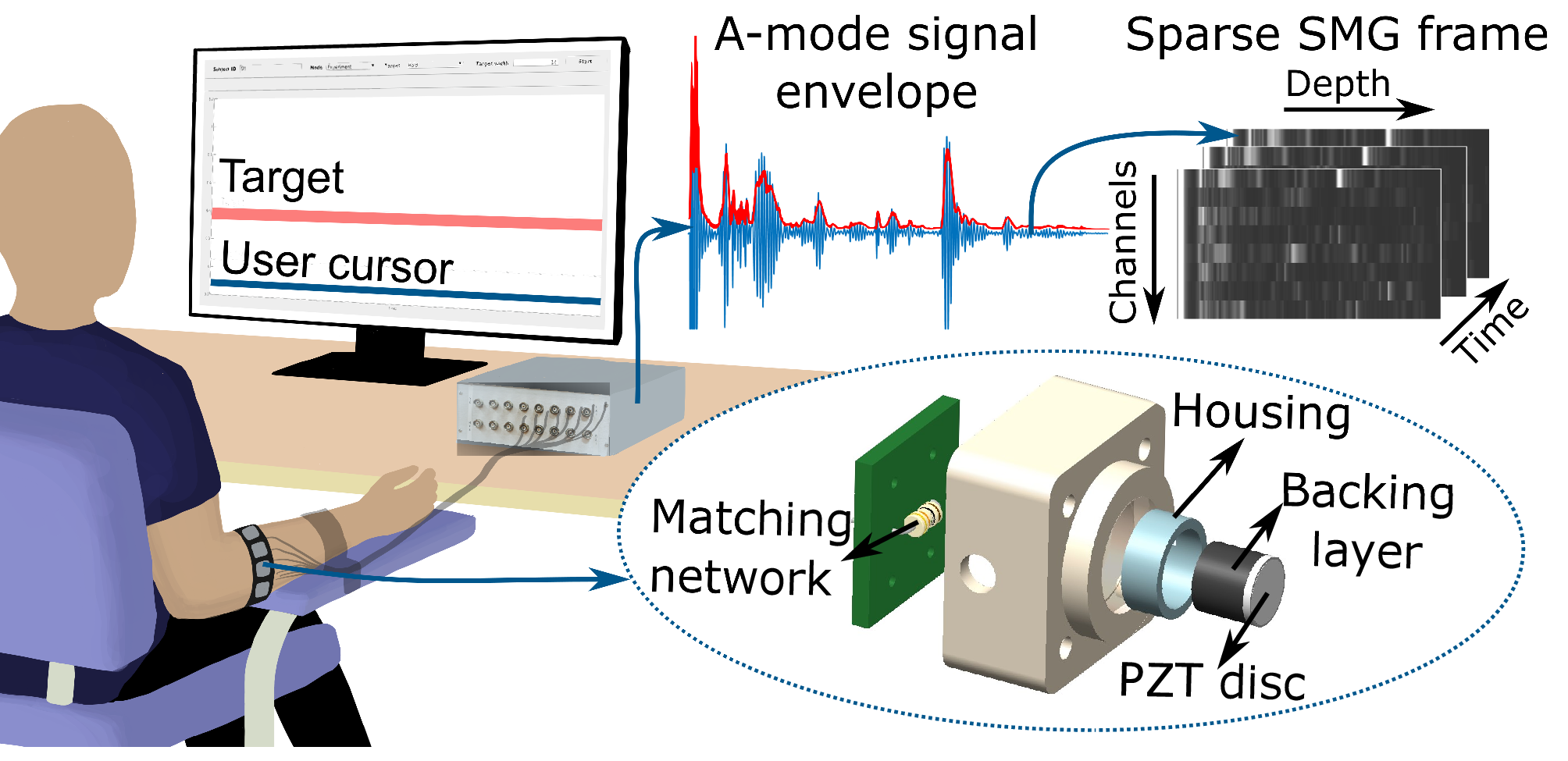}%
\end{wrapfigure}%
\begin{abstract}
Sonomyography (SMG) is a non-invasive technique that uses ultrasound imaging to detect the dynamic activity of muscles. Wearable SMG systems have recently gained popularity due to their potential as human-computer interfaces for their superior performance compared to conventional methods. This paper demonstrates real-time positional proportional control of multiple gestures using a multiplexed 8-channel wearable SMG system. The amplitude-mode ultrasound signals from the SMG system were utilized to detect muscle activity from the forearm of 8 healthy individuals. The derived signals were used to control the on-screen movement of the cursor. A target achievement task was performed to analyze the performance of our SMG-based human-machine interface. Our wearable SMG system provided accurate, stable, and intuitive control in real-time by achieving an average success rate greater than 80\,\% with all gestures. Furthermore, the wearable SMG system's abilities to detect volitional movement and decode movement kinematic information from SMG trajectories using standard performance metrics were evaluated. Our results provide insights to validate SMG as an intuitive human-machine interface. 
\end{abstract}
\end{minipage}}}
\maketitle

\section{Introduction}
Intuitive control of powered upper extremity prosthetics and assistive devices is essential to restore function in individuals with motor disabilities. Detection of volitional motion intent is critical to achieving accurate and intuitive control. Surface electromyography (EMG), a technique that senses the electrical activity in the muscles, is widely used to detect motor intent. Modern EMG-based control strategies provide control over multiple degrees of freedom (DOFs) utilizing measurements performed across multiple channels, referred to as pattern-recognition systems~\cite{Hahne2018, Jiang2012a}. Despite advances in control strategies, only a few DOFs of the myoelectric prostheses could be controlled during real-life use~\cite{Resnik2018}, primarily due to the limitations of EMG. Poor spatial specificity, muscle cross-talk in EMG measurements, electrode shifts, limb position, and force variations result in poor functional outcomes during real-life usage~\cite{Jiang4663628, Scheme2011, Vinjamuri4463048}. Therefore, exploring alternate non-invasive sensing modalities that could robustly detect the volitional movement from muscles is necessary to enable dexterous control of the biomechatronic systems. 

Sonomyography (SMG) has recently gained importance because of its ability to sense voluntary movement intention, which is crucial for efficient biomechatronic control. SMG detects mechanical deformations that occur alongside the electrical potential changes of the muscles during voluntary movement. SMG uses ultrasound imaging techniques to sense the anatomical deformations during dynamic activity, unlike EMG, which senses changes in electrical potentials. Brightness mode (B-mode) ultrasound imaging technique provides spatially resolved images of muscles and their anatomical details. Several studies quantified the changes in anatomical structures such as muscle fiber length~\cite{Zhang9160868}, cross-section~\cite{Hallock9224391}, and pennation angle~\cite{Shi2007e}. They found a high correlation with joint angles, muscle force, and muscle activation levels~\cite{Hodges_2003}, which can be used for human-machine interfaces. B-mode SMG has been widely used to demonstrate gesture classification~\cite{McIntosh2017}. Akhlagi et al. demonstrated the classification of up to 15 gestures with 92\,\% real-time classification accuracy~\cite{Akhlaghi7320970}. SMG has also been used to achieve proportional control~\cite{Shi_2010, Dhawan2019} and perform functional tasks with prostheses~\cite{Engdahl2022a}. Recently, it has also been demonstrated that it is possible to retain high classification accuracy~\cite{Akhlagi8725524} as well as continuous force prediction accuracy~\cite{Kamatham9769757} with only a small subset of the scanlines (4 to 8) chosen from the B-mode images. Therefore, the current research focuses on developing wearable SMG systems that utilize single-element ultrasound sensors to sense muscle activity. 

 Wearable SMG systems consist of a set of single-element ultrasound transducers made of piezoceramic discs~\cite{Yang9185023, Kamatham10101042} or PVDF films~\cite{AlMohimeed_2020, Yan8667816} compared to the large arrays typically found in commercial ultrasound probes. Single-element ultrasound transducers receive echoes reflected by tissues at various depths as one-dimensional amplitude mode (A-mode) signals with the amplitude and depth of the reflected echoes on vertical and horizontal axes, respectively. While B-mode images enable visualization of structural details of the muscles, A-mode signals only provide peaks whenever tissue interfaces are encountered along the path of the transmitted ultrasound waves. Therefore, computational methods relevant to A-mode signals are necessary to detect dynamic muscle activity and quantify biomechanical parameters. Existing algorithms detect muscle activity directly from the A-mode signals or by constructing sparse SMG images. Some algorithms, such as cross-correlation, Pearson's correlation, and peak tracking algorithms, are applicable for both B-mode images~\cite{Shi_2010, Dhawan2019} and A-mode signals~\cite{Jing4575062, Kamatham10176104, Yang8008306}; these algorithms quantify the displacement of the peaks resulting from tissue boundaries to quantify joint angles and muscle forces. However, the popular approach is to extract amplitude, and frequency domain features, namely, mean~\cite{Cai9799771}, standard deviation, root mean square value~\cite{He8477106}, linear fitting coefficients~\cite{Xia8662605, Yang8347147}, frequency spectrum~\cite{Yang9185023}, from A-mode signals and use machine learning techniques to quantify muscle activity.

Wearable SMG systems were also used to demonstrate real-time control of prostheses and virtual tasks successfully. So far, wearable SMG-based has been predominantly focused on gesture classification. Classification of up to 14 finger motions with classification accuracies $>80\%$ has been reported~\cite{Guo9760285, Xia8662605,9872106}. However, a continuous proportional control strategy is critical to achieving natural control of prostheses. The proportional control strategy maps the degree of the user's motor intent to the position, force, or velocity of the prostheses~\cite{Fougner6205630}. This control strategy requires accurate detection of the activation levels of the muscles. However, very few attempts were made to achieve proportional control using wearable SMG systems. Chen et al. first demonstrated proportional control using single-element ultrasound transducers by directly mapping muscle deformation during wrist extension to the opening aperture of the prosthetic hand~\cite{Chen_2010}. In recent years, Yang et al. have implemented proportional pattern recognition to demonstrate the feasibility of classifying gestures along with the contraction levels~\cite{8654210}. However, achieving proportional control from muscle contraction level may lead to fatigue and prolonged use discomfort. Therefore, a real-time proportional position control of multiple gestures has been presented in this paper. This study evaluates the wearable SMG's ability to decode proprioceptive information of various gestures by quantifying muscle deformation. Additionally, movement kinematics and motor control performance were depicted using various performance metrics.

\section{Methods}
\subsection{Wearable sonomyography sensor design}
\begin{figure}[ht]
    \centering
    \includegraphics[width = \columnwidth]{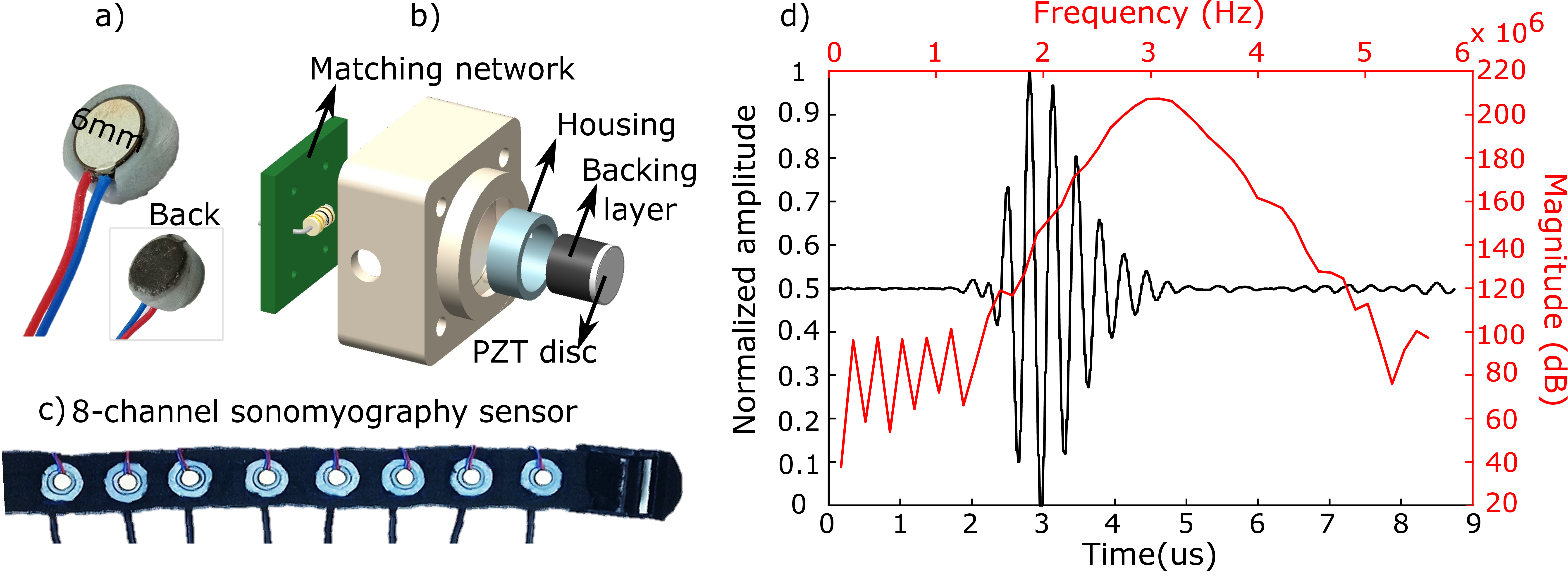}
    \caption{ a) Piezoceramic transducer element with an optimized backing layer, b)  Exploded rendering of the packaged transducer showing the matching network and 3D printed housing, c) 8-channel sensor array on a wearable velcro band, d) Time and frequency domain response from pulse-echo tests of the sensor. }
    \label{fig: sensorDesign}
\end{figure}

An 8-channel wearable SMG sensor was fabricated from Lead Zirconate Titanate (PZT) (SMD063T07R111, Steiner \& Martins Inc, USA) single-element transducers. An acoustic backing layer consisting of 66.7\,\% tungsten powder in an epoxy resin matrix was fabricated following optimization procedures reported in~\cite{Kamatham10101042, Kamatham10176104}. Additionally, an electrical matching network consisting of a parallel L-C tank circuit (L= 5\,$\mu$H and C= 511\, pF) was connected. Each element was characterized by conducting a pulse-echo test as detailed in \cite{Kamatham10101042, Kamatham10176104}. The bandwidth of the transducers was approximately 470$\pm$88\,kHz. Additionally, there was a significant reduction in radial mode vibrations with an improvement of over 100\,dB. Fig.~\ref{fig: sensorDesign} shows the fabricated 8-channel SMG sensor and the time and frequency domain characteristics of a single transducer. The 8-channel array comprises the single sensor units packaged in a customized 3D printed enclosure designed to house the piezoceramic element and the matching network. Connections from each sensor were obtained from 50 ohm RG-174 cables (Make: 8216 0101000, Belden Inc., USA) with an approximate length of 50\,cm. The individual sensor units were attached to a fabric band as shown in Fig.~\ref{fig: sensorDesign}(c) for easy fixation on the forearm. 

\begin{figure*}[ht]
    \centering
    \includegraphics[width = \textwidth]{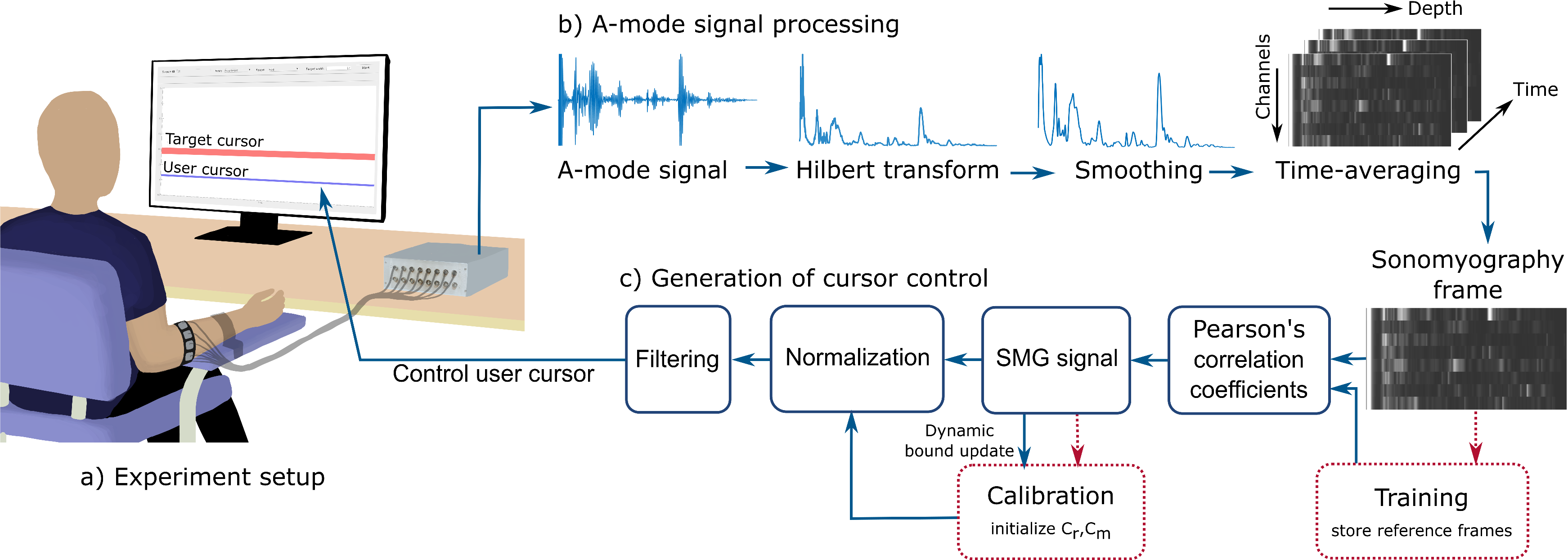}
    \caption{a) Experimental setup: the participants were instrumented with a custom 8-channel SMG sensor connected to a commercial ultrasound pulser-receiver system. The participants performed a real-time target achievement task. b) A-mode signal preprocessing: The preprocessed A-mode signals from all 8 channels were arranged row-wise to form a sparse ultrasound frame. c) Generation of cursor control signal: the control signal was generated by quantifying the gesture position using Pearson's correlation coefficient.}
    \label{fig: Setup}
\end{figure*}

\subsection{Study participants} 
Eight able-bodied individuals participated in the study ($29\pm4$). Participants were informed of the protocols and gave their informed consent to participate in the study. The Institute Ethics Committee (IEC) at the Indian Institute of Technology Delhi approved the study protocols (IITD IEC no: P021/P050).

\subsection{Experimental setup}\label{sec:experimentProtocol}
 The participants were seated comfortably on a chair with their dominant arm on a handrest. The custom-developed wearable SMG sensor array was placed on the forearm approximately 5\,cm from the elbow~\cite{Kamatham10101042}. The wearable  SMG sensor array was connected to a commercial 8-channel multiplexed ultrasound pulser-receiver system (Leceour 8-channel US-MUX, Lecoeur Electronique, France). The US-MUX was operated in pulse-echo mode to allow transmission and reception of ultrasound signals from a single ultrasound transducer. The high-voltage pulser was configured to excite the ultrasound transducers with a unipolar pulse amplitude of 90\,V and pulse width of 49.25\,ns at a pulse repetition frequency (PRF) of 15\,kHz to allow an imaging depth of 10\,cm. The pulse width was optimized by performing an acoustic reflection test to maximize the echo amplitude~\cite{Kamatham10101042}. The received ultrasound echo signals were amplified by a time gain compensation (TGC) amplifier. The gain of the TGC varied linearly from 0\,dB to 50\,dB across the entire depth to compensate for acoustic attenuation in tissue. The amplified signal was then digitized at a sampling rate of 80\,MHz. The RF echo signal was streamed to a PC (Intel Core i7-4790K, 32GB RAM, 2GB NVIDIA GeForce GTX 760) by a USB connection and processed in a custom-developed MATLAB (Version 2022b, Mathworks Inc., Natick MA, USA) software. The MATLAB-based A-mode signal acquisition resulted in a frame rate of 22 frames per second (fps) from all 8 channels. The preprocessing steps are elaborated in the following section. 

\subsection {A-mode signal pre-processing}
The received RF A-mode signal has a signal length of 4000 points. Firstly, the envelope of the RF echo was extracted by computing the analytic signal using the Hilbert transform over a window size of 1000 samples. \hl{} The envelope was then smoothed with a moving average filter having a window size of 25 samples, as shown in Fig.~\ref{fig: Setup}. The smoothed envelopes of all eight channels were then concatenated to form a sparse ultrasound frame of $8\times4000$. The sparse ultrasound frame was further temporally smoothed by averaging three consecutive frames. This time-averaged sparse ultrasound frame was used to generate the SMG signal, as explained in the following section. 

The target achievement task was designed using MATLAB AppDesigner and was displayed on the computer screen. Each experiment session consists of three parts, namely, 1) training, 2) calibration, and 3) target achievement tasks.

\subsubsection{Training phase} During training, a text cue was displayed on the computer screen instructing the participants to rest or perform a specific gesture for 30\,s. The A-mode signals corresponding to the rest and maximum positions of the specific gesture were preprocessed and saved as rest reference and motion reference ultrasound frames, respectively. Pearson's 2D correlation coefficients were calculated between the incoming time averaged ultrasound frame and reference frames corresponding to rest and maximum motion state (see Section~\ref{sec:experimentProtocol}). The SMG signal was calculated as, 
    \begin{equation}
     \label{Eq:1}
       S = \frac{(1- C_r)}{(1-C_r)+(1-C_m)}
     \end{equation}
$C_r$ is the computed correlation coefficient of the incoming frame with the rest reference frame, and $C_m$ is the computed correlation coefficient of the incoming frame with the motion reference frame.
    
\subsubsection{Calibration phase} During calibration, the participants were instructed to perform the selected gesture and rest thrice for 10\,s each. The SMG signal ($S$) was computed based on the reference frames obtained during the training phase. The lower and upper bounds of the SMG signal were captured during the 10s periods and were subsequently used to normalize the SMG signal ($S$) to a range of [0 1]. Therefore, when the participants were fully relaxed, the normalized SMG signal would be '0', and when the participant completed the gesture, the signal would be '1'. The normalized SMG signal value would proportionally scale between '0' and '1' when the user performed the gesture partially. 

\subsection{Experiment 1: Optimizing sensor placement across gestures}

An unconstrained proportional control task was performed to determine the optimal position on the forearm for placement of the sensor array and to derive the relationship between the SMG signal and joint angles for each gesture. Five out of eight participants were recruited for this pilot experiment. The participants wore the wearable SMG sensor as described in Section~\ref{sec:experimentProtocol}. A data glove (5DT Data Glove Ultra, Fifth Dimension Technologies, USA) measured the metacarpophalangeal joint (MCP) angle for each finger. On-screen visual cues were displayed during the task, instructing the participants to perform the motion or rest. The full-range gesture was performed for a duration of 15\,s during motion phases. The duration when from rest state to reach the maximum range of motion for each gesture is termed as upslope. whereas returning from maximum position to rest is termed as downslope. Each gesture was repeated five times, interleaved with 10\,s of resting phases. The participants performed three gestures: power grasp (PG), tripod grasp (TG), and index point (IP). This was repeated by shifting the SMG sensor 5\,cm distal to the initial location. The SMG signal and the data glove signals were saved for offline analysis. The SMG signals obtained during the motion phases were computed offline, and a linear fit was considered to assess the relationship between SMG signals and gesture-specific joint angles.

\subsection{Experiment 2: Target achievement task} The task consists of an on-screen virtual cursor controlled by the normalized SMG signal. A virtual target was presented at levels 0.2, 0.4, 0.6, 0.8, and 1 in randomized order interleaved with 15\,s of rest. The user was instructed via visual cues to attain the target by performing the specific gesture proportionally. On presentation of the target, the user was instructed to reach the target and stay within its bounds for a continuous period of at least 1.5\,s for the target to be successfully acquired. The user was provided a maximum timeout period of 15\,s to attempt to acquire the target. The thickness of the target, or target width (W), was varied to modulate the difficulty of the task. The experiment was performed with three target widths of 5\,\%, 10\,\%, and 15\,\%. Four different gestures, namely, power grasp (PG), index point (IP), tripod grasp (TG), and wrist rotation (WR), were performed. Each test condition was repeated for three trials. The user-generated cursor trajectory for each trial was stored for further analysis.

\subsection{Performance metrics}
\begin{figure} [!ht]
    \centering
    \includegraphics[width = \columnwidth]{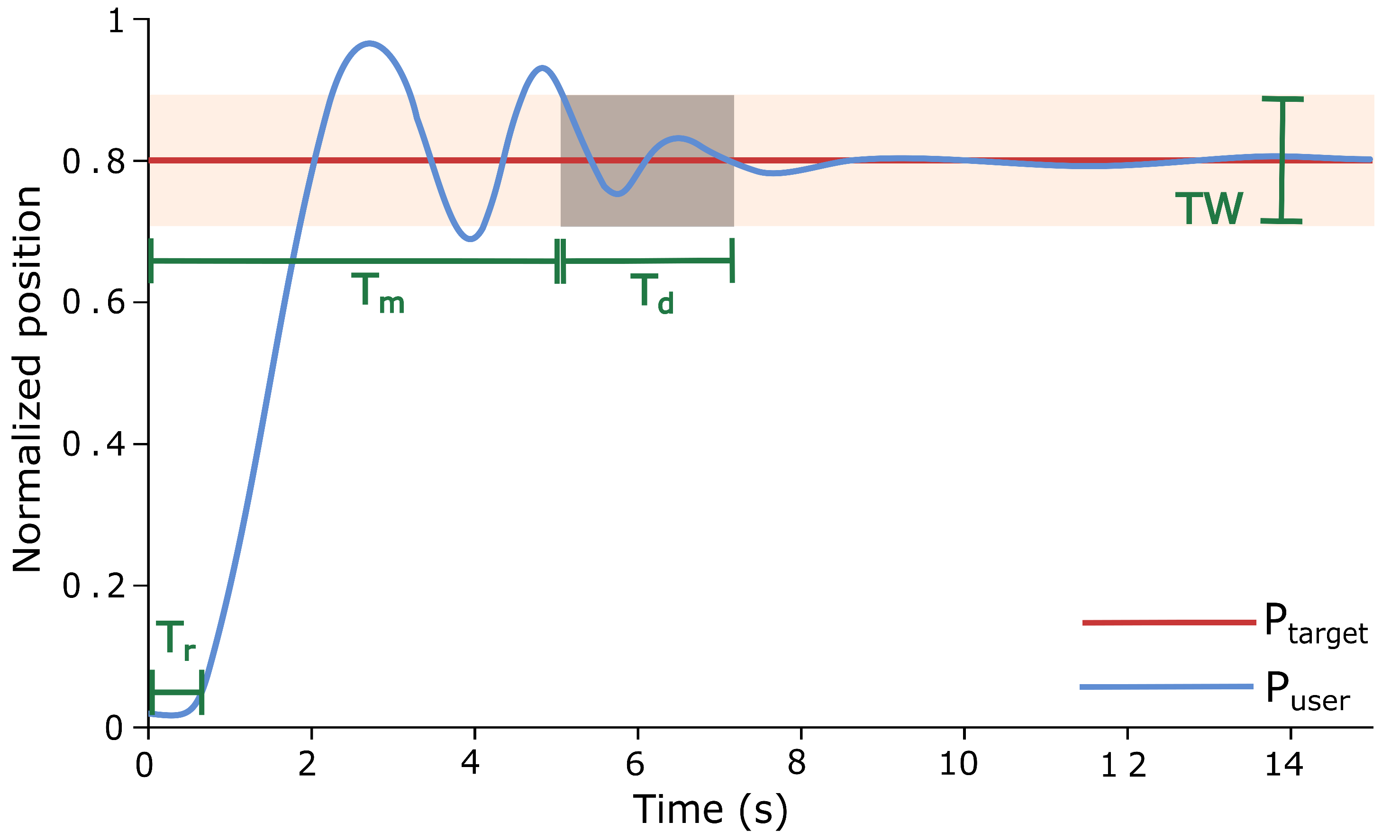}
    \caption{Example of user trajectory and presented target. The outcome metrics, namely, movement time ($T_m$), endpoint error, endpoint stability, path efficiency, and maximum velocity, were calculated from user trajectory.}
    \label{fig: outcome_metrics}
\end{figure}

The task performance of the participants in experiment 2 was evaluated using the metrics extracted from the cursor trajectory as shown in Fig~\ref{fig: outcome_metrics}.
\subsubsection{Success rate} 
The success rate indicates the proportion of the total trials executed successfully by the user. A trial is a success if the participant reaches the target and stays within the target width for at least 1.5\,s of dwell time ($T_d$) as shown in Fig.~\ref{fig: outcome_metrics}. The success rate is calculated as,
\begin{equation}
    Success\,rate = \frac{Number\ of\ successful\ targets}{Number\ of\ targets\ presented} \times 100 \%
\end{equation}
\subsubsection{Movement time}
Movement time ($T_m$) is the time elapsed between the presentation of the target and time instant when the user successfully remained within the target width (dwell window) for a contiguous period defined by the dwell time ($T_d$) as shown in Fig.~\ref{fig: outcome_metrics}. Fitt's law predicts that the movement time increases linearly with task difficulty for goal-directed human-machine interactions~\cite{Fitts1992_movement}. The difficulty of the task (ID) is modulated by the distance of the target from the starting position ($D$), as well as the width of the target ($W$), and is defined as,
\begin{equation}
    ID = log_{2}\Bigg[ 1+\frac{D}{W}\Bigg]
\end{equation}
A linear fit was obtained between the index of difficulty and movement time. The Fitt's throughput was estimated as the inverse of the slope of the best-fit line, and the y-axis intercept was obtained as an estimate of the reaction time for each grasp type. 

\subsubsection{Endpoint error}
Endpoint error ($E_{error}$) provides a measure of the average deviation of the user cursor and the intended target once the target has been successfully acquired, as shown in Fig.~\ref{fig: outcome_metrics}. It is defined as the difference between the target position and the mean value of the participant trajectory within the dwell time window. It is calculated as,
\begin{equation}
    E_{error} = \sum_{t=T_{m}}^{T_{m} + T_{d}} \frac{P_{target} (t) - P_{user} (t)} {N} 
\end{equation}
Here, $N$ is the number of samples of $P_{user}(t)$, $T_{m}<t<T_{m} + T_{d}$, $P_{target}(t)$ is the position of the target presented to the user at time, $t$ and $P_{user} (t)$ is the position of the user's cursor at time, $t$.

\subsubsection{Endpoint stability}
Endpoint stability ($E_{stability}$) provides a measure of the jitter of the user's cursor once the target has been successfully acquired, as shown in Fig.~\ref{fig: outcome_metrics}. It is defined as the standard deviation of the participant trajectory within the dwell time window. It is calculated as,
\begin{equation}
    E_{stability} = \sum_{t=T_{m}}^{T_{m} + T_{d}} \frac{(P_{user} (t) - \mu_{user})^2}{N}
\end{equation}
Here, $N$ is the number of samples of $P_{user}(t)$, $T_{m}<t<T_{m} + T_{d}$ and, $\mu_{user}$ is defined as in Eq.~\ref{eq:mu_user},
\begin{equation}
\label{eq:mu_user}
    \mu_{user} = \sum_{t=T_{m}}^{T_{m} + T_{d}} \frac{P_{user} (t)}{N}
\end{equation}

\subsubsection{Path efficiency}
Path efficiency ($\eta$) is a measure of the total path length of the user's trajectory compared to an idealized trajectory defined by instantaneous target acquisition at $t=0$ as in eq.~\ref{eq:path_eff}:
\begin{equation}
\label{eq:path_eff}
    \eta = \frac{P_{target}-P_{user}(0)}{\sum_{t=0}^{T_{m}}\lvert P_{user}(t+1) - P_{user} (t) \rvert} \times 100 \%
\end{equation}

\subsubsection{Maximum velocity} The velocity profile of precision dexterous grasps is known to scale with object size, as well as the distance to the object in healthy adults~\cite{Hoff1993, MonWilliams2001}, and could provide valuable insights into the motor behavior experienced while controlling the SMG-based muscle computer interface. Hence, the maximum velocity of the user-generated trajectory for each target position was calculated as,
\begin{equation}
    v_{max} = max \Bigg\lvert \frac{d P_{user} (t)}{dt}\Bigg\rvert, 0\leq t \leq T_{m} 
\end{equation}

\subsection{Statistical tests}
A two-way repeated measures ANOVA was used to determine whether the interaction and main effects of target position and width are statistically significant. The Greenhouse-Geisser correction was applied for conditions where Mauchly's sphericity test was violated. IBM SPSS Statistics (Version 28.0, IBM Corp, Armonk, NY, USA) was used for all statistical analyses.

\begin{figure}[!t]
    \centering
    \includegraphics[width = \columnwidth]{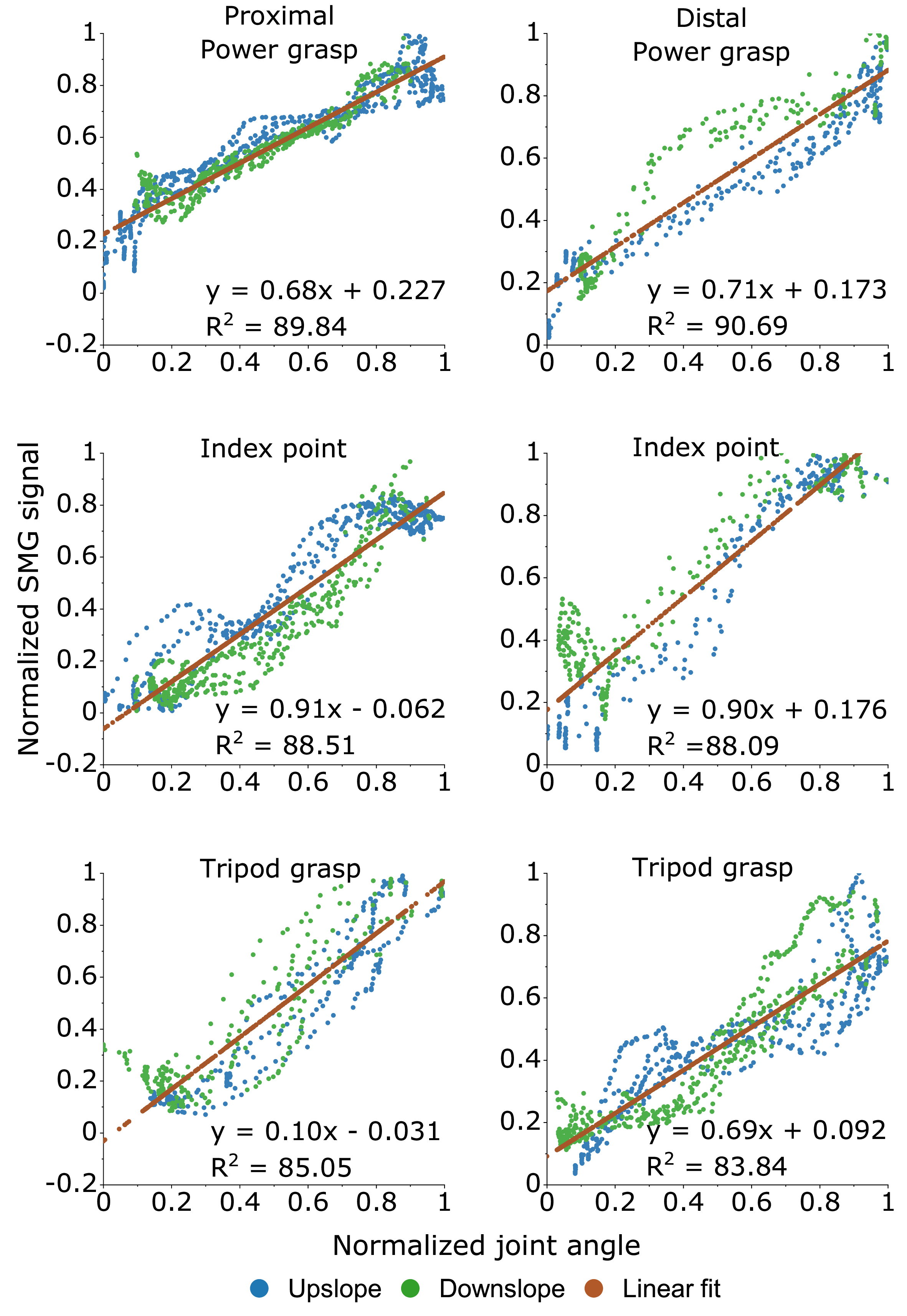}
    \caption{Relationship between MCP joint angle and SMG signal for three gestures with the sensor array placed at proximal and distal locations on the forearm. Plots show the best $R^2$ values obtained for each gesture at each sensor position.}
    \label{fig: DGVsSMG}
\end{figure}

\section{Results}

\subsection{Experiment 1: Optimizing sensor placement across gestures}

Fig.~\ref{fig: DGVsSMG} shows the relationship between normalized joint angle and the SMG signal for three gestures, PG, IP, and TG, for proximal and distal placement of wearable SMG sensor. The best case $R^2$ values obtained for each condition were considered for Fig.~\ref{fig: DGVsSMG}. The mean and standard deviation in $R^2$ values for all participants are listed in Table.~\ref{meanR2values}. A two-sample t-test was performed to determine whether the difference in linearity obtained at the two sensor locations was statistically significant. Results demonstrated that proximal placement of the sensors on the forearm resulted in significantly higher linearity for TG (t(4) = 4.241, p = 0.013). For PG, proximal placement led to higher linearity than distal sensor placement. However, this difference in linearity was not found to be statistically significant for PG  (t(4) = 1.934, p = 0.125). Similarly, for IP, the difference in linearity between the two positions was not found to be statistically significant (t(4) = -2.194, p = 0.093). Therefore, subsequent experiments were performed with the sensor array placed at the proximal location to improve the congruence of the SMG signal with the MCP joint angles.  
\renewcommand{\arraystretch}{2}
\begin{table}[ht]
\caption{Mean and standard deviation of $R^2$ values obtained for each gesture at proximal and distal positions}
\centering
\begin{adjustbox}{width=\columnwidth}
\begin{tabular}{ccccccc}\toprule
& \multicolumn{3}{c}{{\Large Proximal}}&\multicolumn{3}{c}{{\Large Distal}}
\\\cmidrule(lr){2-4}\cmidrule(lr){5-7}
\multirow{1}{*}{}  
             & {\Large Power grasp}             
             & {\Large Index point}         
             & {\Large Tripod grasp}          
             & {\Large Power grasp}             
             & {\Large Index point}          
             & {\Large Tripod grasp}      \\
             \midrule
\multirow{1}{*}{{\Large Mean}}             
            & {\Large $85.89$}
            & {\Large $71.24$}
            & {\Large $78.22$}
            & {\Large $78.67$}
            & {\Large $81.22$}
            & {\Large $70.55$} \\
\multirow{1}{*}{{\Large (Standard deviation)}}  
            & {\Large ($\pm 2.959$)}
            & {\Large ($\pm 17.212$)}
            & {\Large ($\pm 6.870$)}
            & {\Large ($\pm 9.819$)}
            & {\Large ($\pm 9.272$)}
            & {\Large ($\pm 9.803$)} \\
            \bottomrule

\end{tabular}
\label{meanR2values}
\end{adjustbox}
\end{table}

\subsection{Experiment 2: Target achievement task}

\begin{figure} [ht]
    \centering
    \includegraphics[width = \columnwidth]{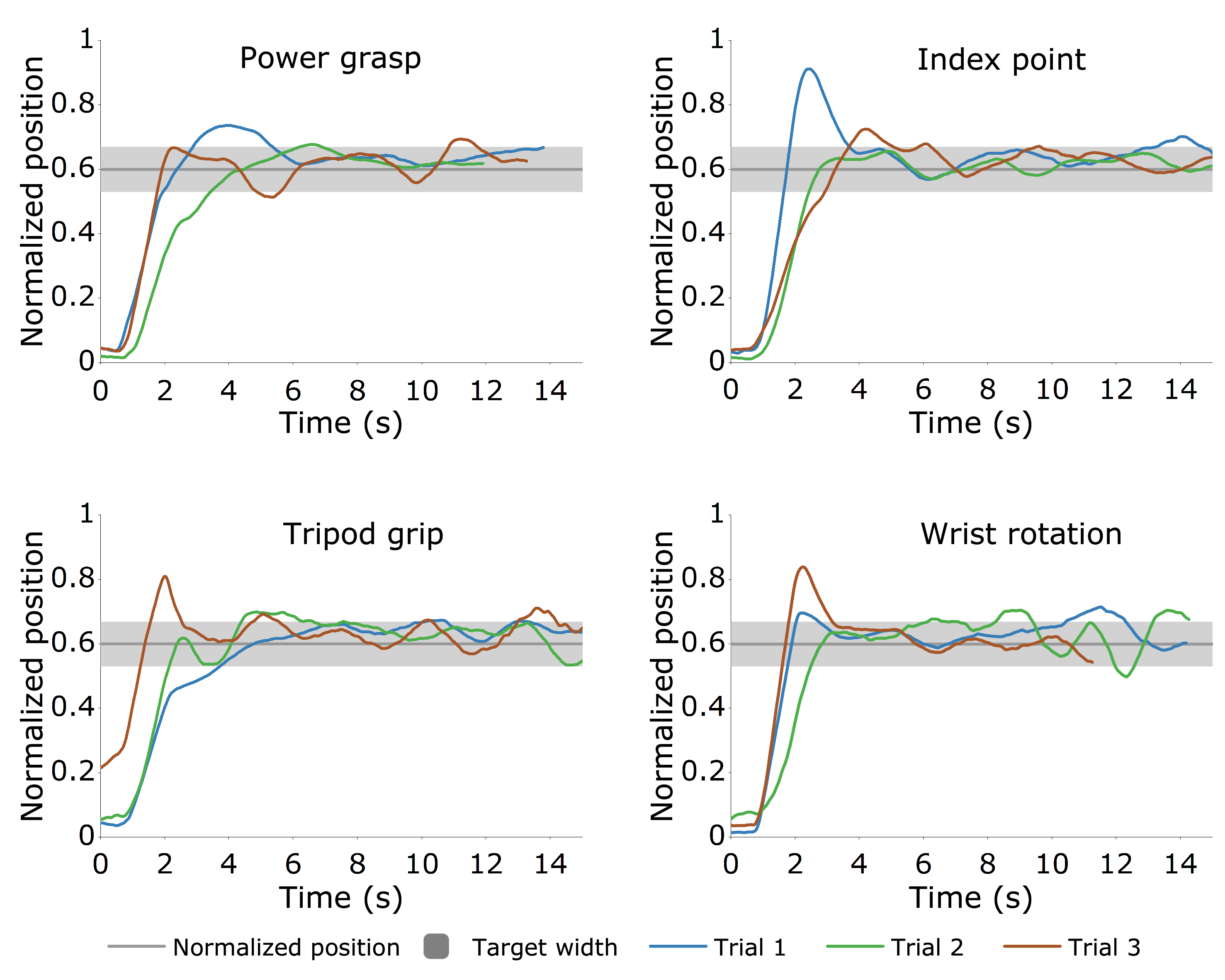}
    \caption{The movement trajectories of a representative participant for a target presented at 0.6 achieved using all gestures}
    \label{fig: user_trajectories}
\end{figure}

Fig.~\ref{fig: user_trajectories} shows the SMG signal trajectories of a representative participant for a target presented at 0.6. The participants performed three trials for each target position. The figure shows the trajectories attained during each trial for all four grasps. 

\subsubsection{Success rate}
\begin{figure}[!bh]
    \centering
    \includegraphics[width = \columnwidth]{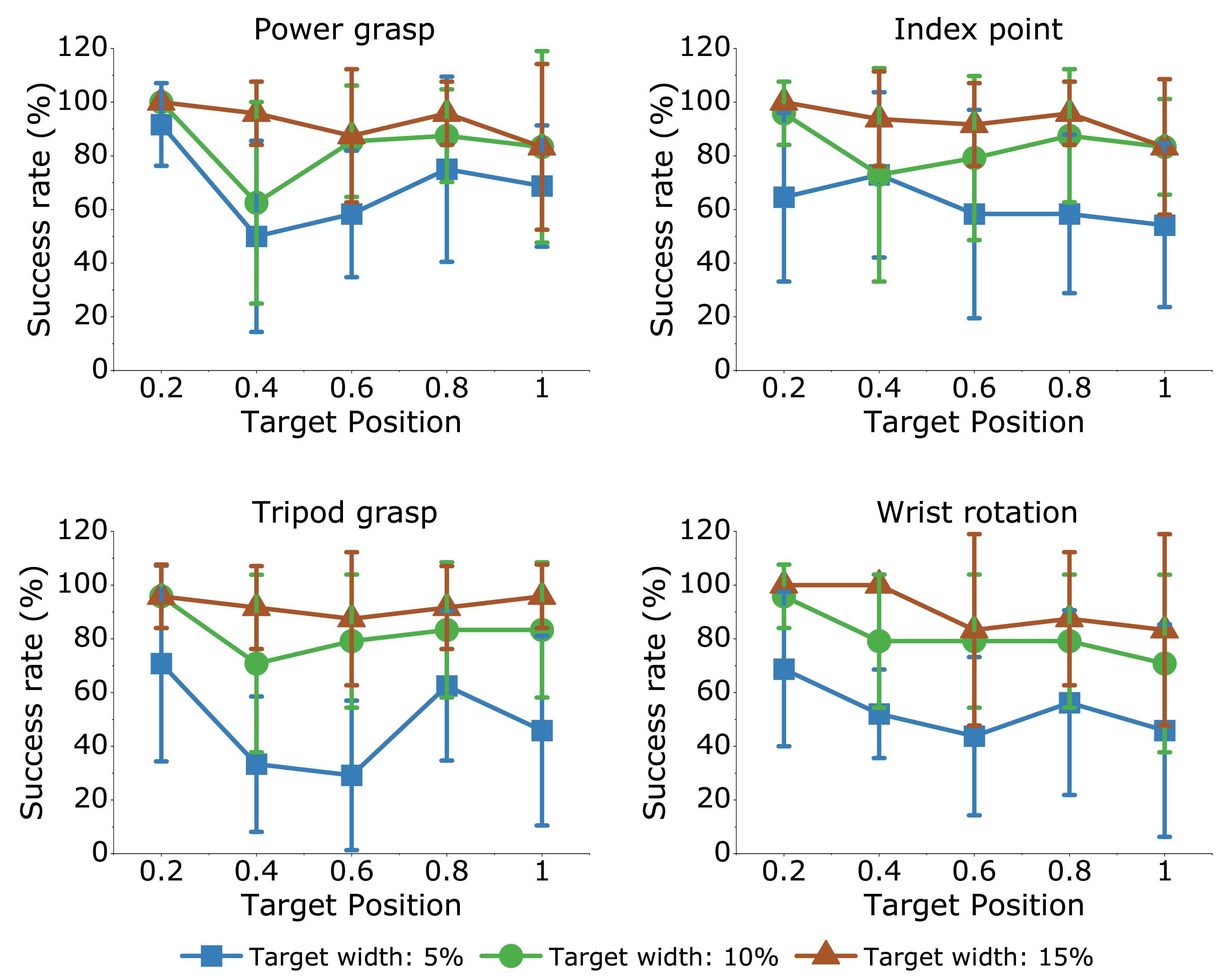}
    \caption{Success rates achieved with different gestures at each position for three target widths. The participants achieved higher success rates target width of 15\%.}
    \label{fig:successrate}
\end{figure}

Fig.~\ref{fig:successrate} shows the success rates achieved at all target positions for three different target widths for all the gestures. The participants achieved average success rates of $76.6(\pm16.96$)\%,  $73.6(\pm18.69$)\%, $68.7(\pm24.87$)\%, $69.8(\pm21.47$)\% for PG, IP, TG, and wrist rotation respectively. It was observed that participants successfully acquired a larger proportion of targets closer to rest (0.2) compared to other targets located farther away. However, this effect of target position on success rate was only significant for PG  ($F(4,28) = 4.20, p=0.009$)  and TG ($F(4,28) = 4.680, p = 0.005$). There was a significant ($p<0.001$) improvement in the success rates when the target width was increased for all the gestures. With a target width of $15\%$, success rates as high as $100\%$ were achieved. There was no significant difference in the success rate between the gestures for a given target width, indicating that target width modulates task difficulty across all gestures ($p=0.101$ for 5\%, $p=0.898$ for 10\%, and $p=0.682$ for 15\%). 
\renewcommand{\arraystretch}{1.5}
 
\begin{table*}[!ht]
\caption{p-values obtained to determine the statistically significant effects of target position and target width on various performance metrics for all gestures}
\centering
\begin{adjustbox}{width=\textwidth}
\fontsize{12pt}{12pt}
\begin{tabular}{ccccccccc}\toprule
& \multicolumn{2}{c}{{\large Power grasp}}&\multicolumn{2}{c}{{ \large Index point}}&\multicolumn{2}{c}{{\large Tripod grasp}}&\multicolumn{2}{c}{{\large Wrist rotation}}
\\\cmidrule(lr){2-3}\cmidrule(lr){4-5}\cmidrule(lr){6-7}\cmidrule(lr){8-9}
\multirow{1}{*}{}  
             & \large Target Position             
             & \large Target width           
             & \large Target Position             
             & \large Target width             
             & \large Target Position             
             & \large Target width           
             & \large Target Position             
             & \large Target width        \\
             \midrule
\multirow{2}{*}{\large Success rate  }             
            & $\textbf{0.009}^*$
            & $\textbf{0.002}^*$
            & $0.525$
            & $<\textbf{0.001}^*$
            & $\textbf{0.005}^*$
            & $<\textbf{0.001}^*$
            & $0.075$
            & $<\textbf{0.001}^{*,\dagger}$ \\
\multirow{1}{*}{}  
            & \small $F(4,28) = 4.20$
            & \small $F(2,14) = 9.99$
            & \small $F(4,28)=0.82$
            & \small $F(2,14)=21.36$
            & \small $F(4,28) = 4.68$
            & \small $F(2,14)=20.50$
            & \small $F(4,28) = 2.38$
            & \small $F(1.183,8.284) = 24.84$ \\
            \bottomrule
\multirow{2}{*}{\large Movement time} 
            & $\textbf{0.003}^*$        
            & $\textbf{0.002}^*$ 
            & $<\textbf{0.001}^*$      
            & $\textbf{0.007}^*$
            & $0.150$   
            & $\textbf{0.004}^*$
            & $\textbf{0.008}^*$   
            & $\textbf{0.016}^{*,\dagger}$ \\
\multirow{1}{*}{}  
            & \small $F(4,44) = 4.82$
            & \small $F(2,22) = 8.43$
            & \small $F(4,60) = 6.45$
            & \small $F(2,30) = 5.97$
            & \small $F(4,24) = 1.86$
            & \small $F(2,12) = 9.26$
            & \small $F(4,36) = 4.05$
            & \small $F(1.127,10.146)=7.88$ \\
            \midrule
\multirow{2}{*}{\large Endpoint error} 
            & $<\textbf{0.001}^*$          
            & $0.310^\dagger$     
            & $<\textbf{0.001}^*$       
            & $0.110$
            & $\textbf{0.022}^*$     
            & $<\textbf{0.001}^*$             
            & $\textbf{0.012}^*$     
            & $0.099^\dagger$\\
\multirow{1}{*}{} 
            & \small $F(4,44)=7.90$
            & \small $F(1.337,14.703) = 1.19$
            & \small $F(4,60) = 14.12$
            & \small $F(2,30) = 2.37$
            & \small $F(4,24) = 3.51$
            & \small $F(2,12)  = 13.96$
            & \small $F(4,36) = 3.77$
            & \small $F(1.162,10.462) = 3.20$ \\
            \midrule
             
\multirow{2}{*}{\large Endpoint stability} 
            & $<\textbf{0.001}^*$        
            & $<\textbf{0.001}^*$ 
            & $<\textbf{0.001}^*$   
            & $<\textbf{0.001}^*$
            & $\textbf{0.031}^{*,\dagger}$      
            & $<\textbf{0.001}^*$
            & $<\textbf{0.001}^*$     
            & $<\textbf{0.001}^*$\\
\multirow{1}{*}{}     
            & \small $F(4,44) = 6.41$
            & \small $F(2,22) =34.95$
            & \small $F(4,60) = 10.33$
            & \small $F(2,30) = 13.24$
            & \small $F(1.836,11.015) = 4.99$
            & \small $F(2,12) = 66.32$
            & \small $F(4,36) = 7.95$
            & \small $F(2,18) = 47.13$ \\
             \midrule
             
\multirow{2}{*}{\large Path efficiency} 
            & $<\textbf{0.001}^*$      
            & $0.303$  
            & $<\textbf{0.001}^*$       
            & $\textbf{0.011}^*$
            & $<\textbf{0.001}^*$   
            & $0.115$
            & $<\textbf{0.001}^*$      
            & $0.104^\dagger$\\
\multirow{1}{*}{}    
            & \small $F(4,44) = 14.26$
            & \small $F(2,22) = 1.26$
            & \small $F(4,60) = 20.26$
            & \small $F(2,30 )= 5.20$
            & \small $F(4,24) = 11.68$
            & \small $F(2,12) = 2.60$
            & \small $F(4,36) = 14.16$
            & \small $F(1.241,11.168) = 3.02$ \\
             \midrule
             
\multirow{2}{*}{\large Maximum velocity} 
            & $<\textbf{0.001}^*$       
            & $0.290$
            & $<\textbf{0.001}^*$    
            & $<\textbf{0.001}^*$
            & $<\textbf{0.001}^*$      
            & $0.290$ 
            & $<\textbf{0.001}^*$    
            & $0.320$ \\
\multirow{1}{*}{}    
            & \small $F(4,44) = 54.28$
            & \small $F(2,22) =1.31$
            & \small $F(4,60) = 110.70$
            & \small $F(2,30) = 9.07$
            & \small $F(4,24) = 11.03$
            & \small $F(2,12) = 1.37$
            & \small $F(4,36) = 87.27$
            & \small $F(2,18) = 1.21$ \\
        \bottomrule\\
      \multicolumn{9}{l}{ {\small *indicates statistically significant p-values, $\dagger$ applied Greenhouse-Geisser correction}}\\
\end{tabular}
\label{tab:p-values}
\end{adjustbox}
\end{table*}
\subsubsection{Movement time}
\begin{figure}[!th]
    \centering
    \includegraphics[width = \columnwidth]{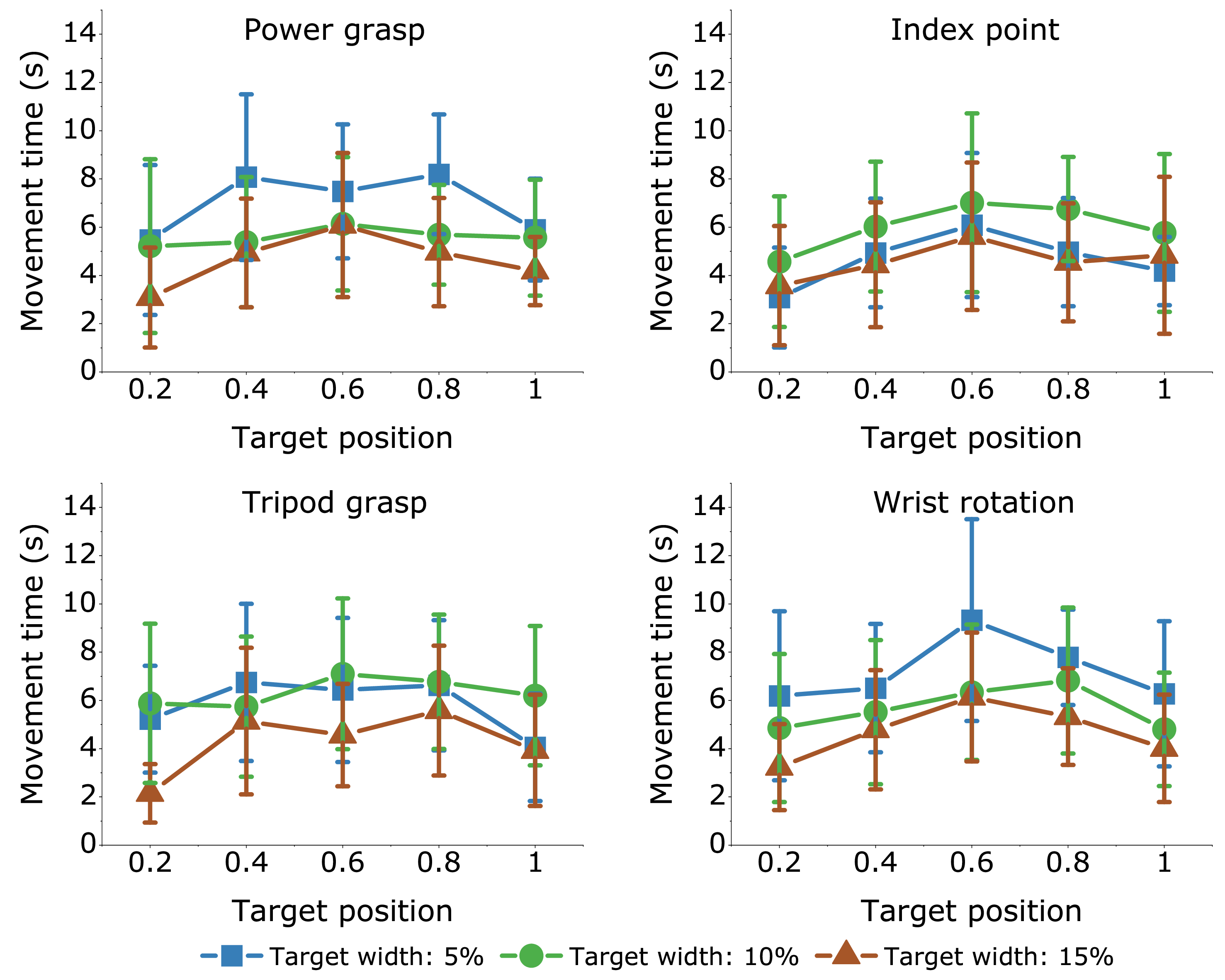}
    \caption{Movement times achieved for different gestures for all target positions and target widths. A significant reduction in movement time can be seen when easier targets (15\% target width) were presented.}
    \label{fig: MovementTime}
\end{figure}

Fig.~\ref{fig: MovementTime} shows the movement times of all the participants for every target position at each target width. An average movement time of $5.6\pm2.9$ s was attained for all four grasps and is not significantly different ($p=0.321$) for different gestures. The movement time for targets at 0.2 and 1 were lower than those at 0.2, 0.4, and 0.6. This effect of target position on movement time was significant in PG  ($F(4,44) = 4.82, p = 0.003$), IP ($F(4,60) = 6.45, p < 0.001$), and wrist rotation  ($F(4,36) = 4.05, p = 0.008$) but not statistically significant for tripod ($F(4,24) = 1.86, p=0.15$) as indicated in Table~\ref{tab:p-values}. However, there was a significant effect of target width on the movement time across all the gestures ( $p < 0.05$) as shown in Table~\ref{tab:p-values}
\subsubsection{Endpoint error and Endpoint stability}
\begin{figure}[!bh]
    \centering
    \includegraphics[width = \columnwidth]{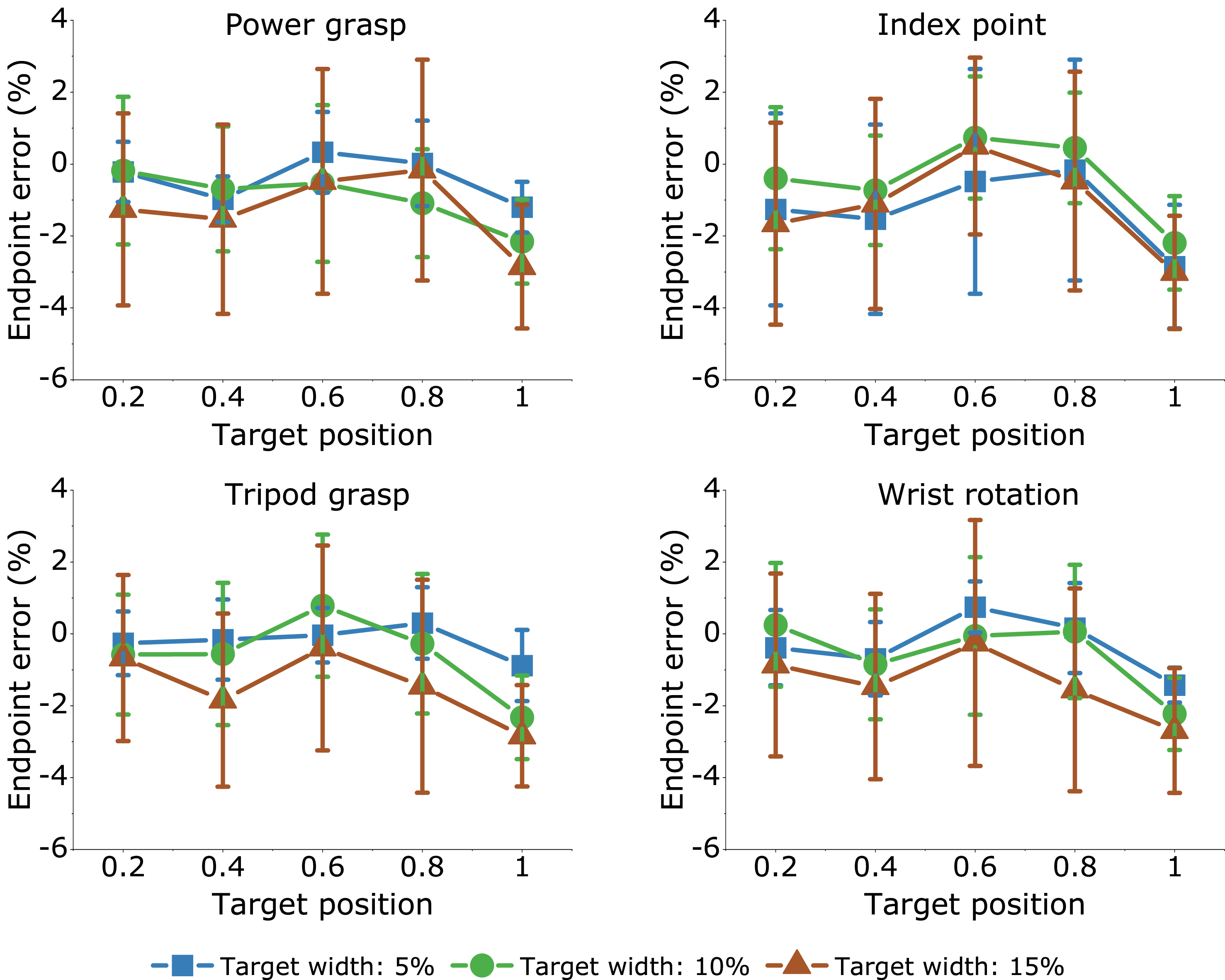}
    \caption{Endpoint errors for each gesture at all target positions and target widths. Endpoint errors were within 5\% of the range.}
    \label{fig:endpointError}
\end{figure}

 Fig.~\ref{fig:endpointError} shows the endpoint error at all target positions and widths in all gestures. For all the gestures, the average position error was $-0.8\pm2 \% $; the negative sign indicates that the user cursor is below the target. The target position significantly affects the endpoint error across all positions ($p<0.001$). Interestingly, the target width had no significant effect on the endpoint error for all gestures except TG ($F(2,12)  = 13.96, p<0.001$). 
 
\begin{figure}[!bh]
    \centering
    \includegraphics[width = \columnwidth]{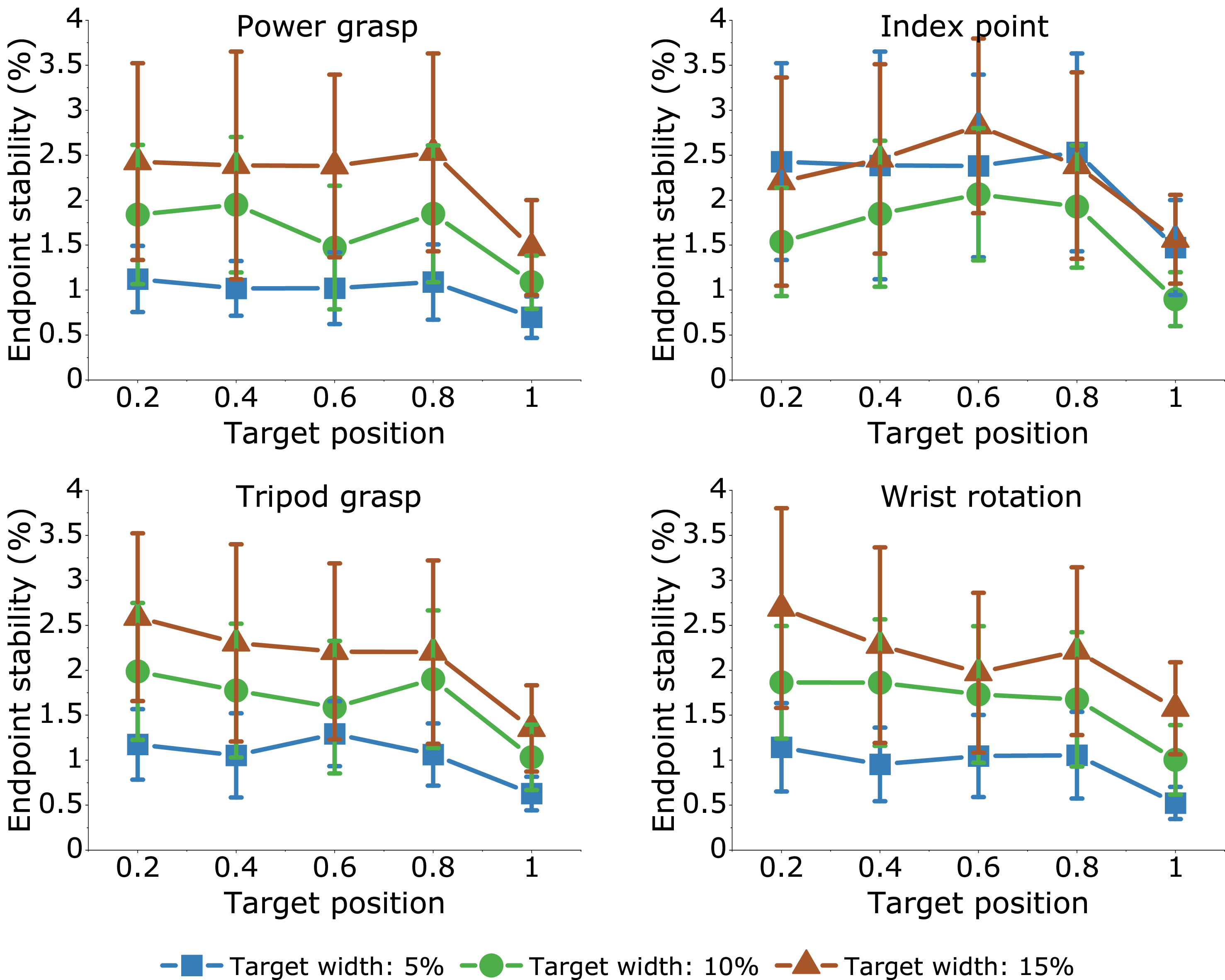}
    \caption{Endpoint stability for each gesture at all target positions and target widths. The standard deviation of the jitter was within 5\% of the range. }
    \label{fig:endpointStability}
\end{figure}

Fig.~\ref{fig:endpointStability} shows the endpoint stability error achieved for all the gestures at all target positions and widths. The average standard deviation in the user trajectory within the dwell time window was $1.7\pm1\%$ for all the gestures. There is a significant ($p<0.05$) effect of target position and width on endpoint stability for all gestures. 

\subsubsection{Path efficiency}
\begin{figure}[!bh]
    \centering
    \includegraphics[width = \columnwidth]{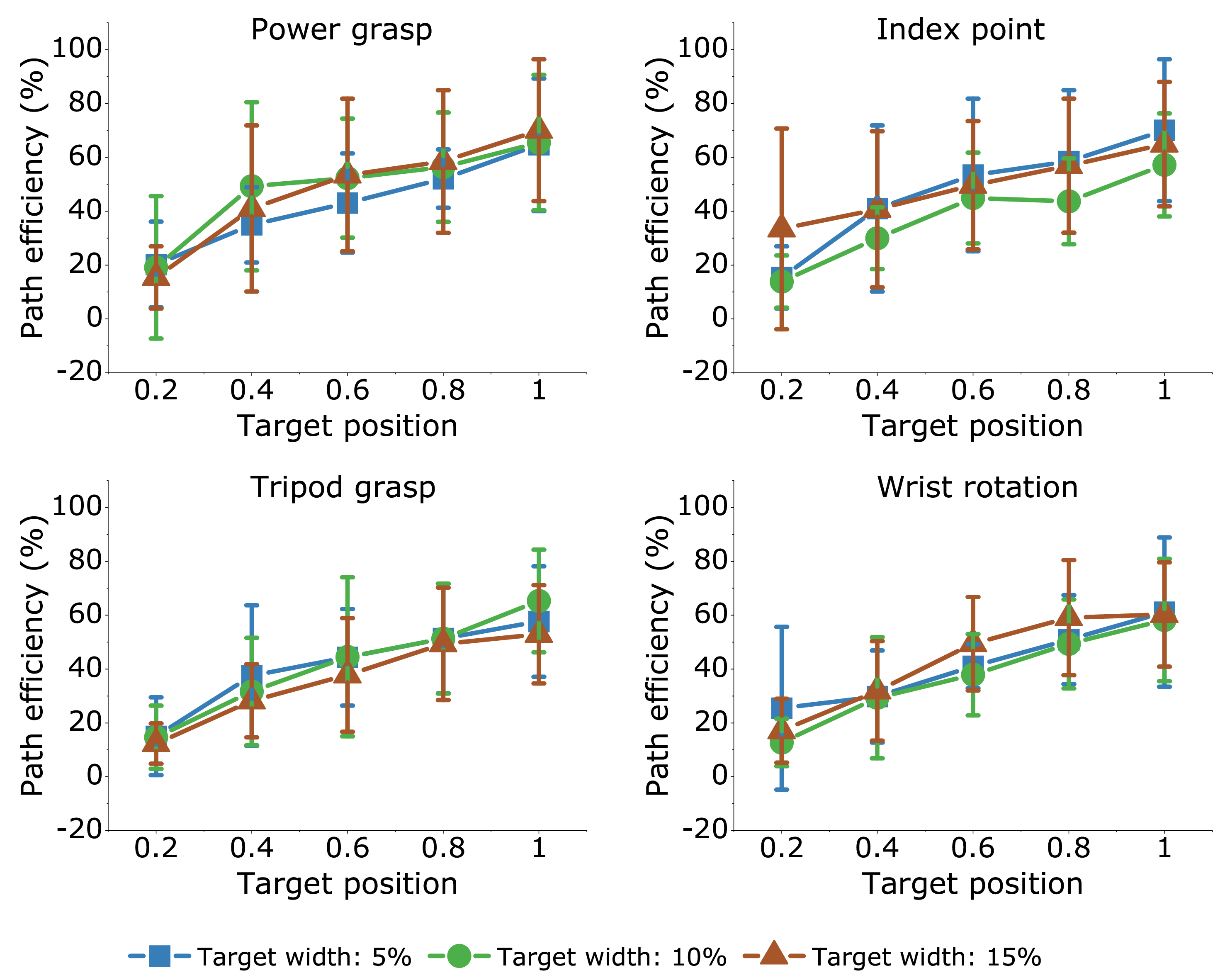}
    \caption{Path efficiency of the user trajectories for all target positions and widths for all gestures.}
    \label{fig: Pathefiiciency}
\end{figure}

Fig.~\ref{fig: Pathefiiciency} shows the path efficiency of the trajectories at all target positions and widths for all the gestures. The path efficiency increased significantly ($p<0.001$) with an increase in the target position. Target width had no significant effect on path efficiency for PG  ($F(2,22) = 1.26, p = 0.303 $), TG ($F(2,12) = 2.60, p < 0.115$), and wrist rotation ($F(1.241,11.168) = 3.02, p < 0.104$). However, target width significantly affected path efficiency for the IP ($F(2,30 )= 5.203, p = 0.011$).

\subsubsection{Maximum velocity}
\begin{figure}[!th]
    \centering
    \includegraphics[width = \columnwidth]{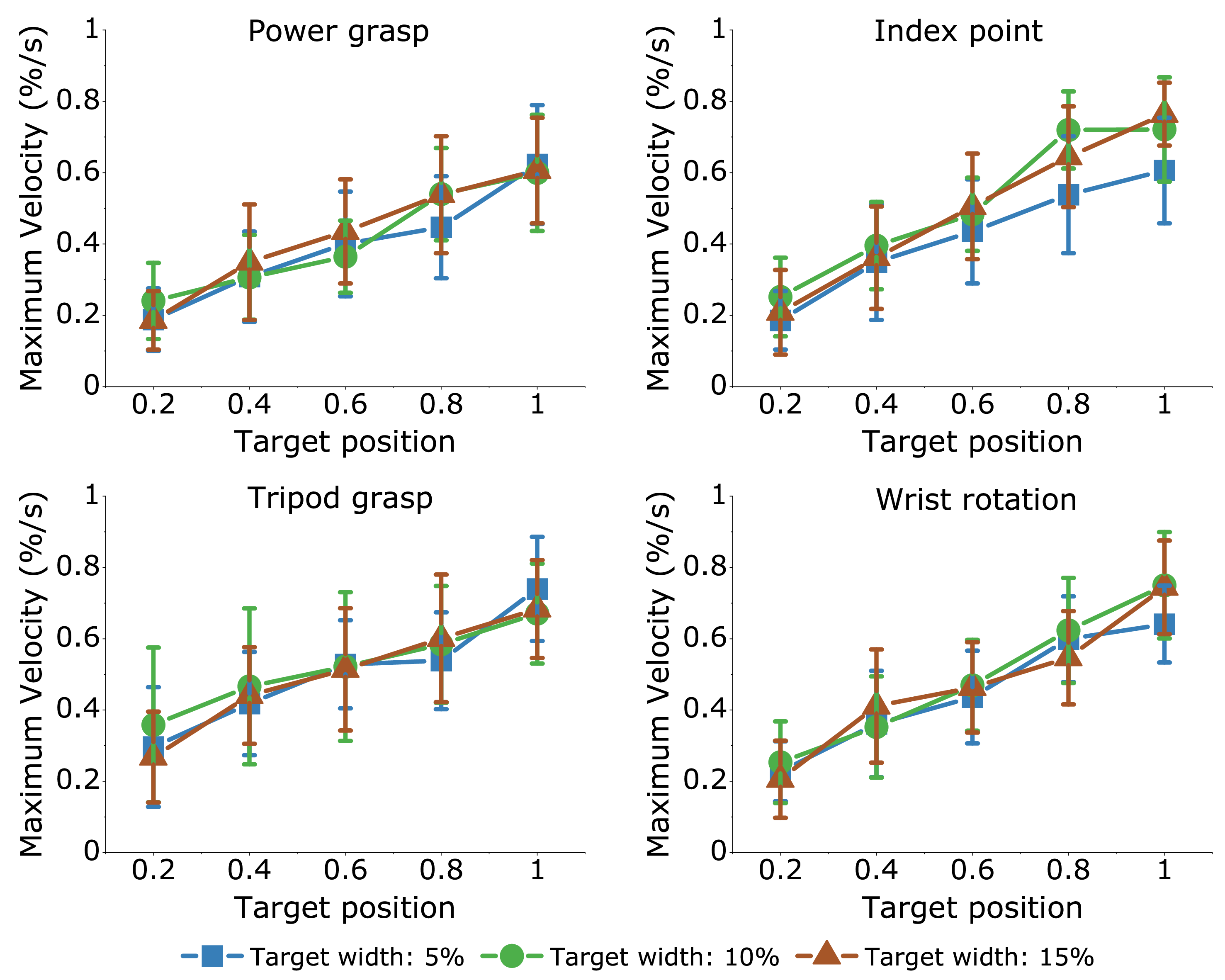}
    \caption{Maximum velocity derived from the user trajectories for all target positions and widths for all gestures. Maximum velocity scaled linearly with target distance.}
    \label{fig: Velocity}
\end{figure}

Fig.~\ref{fig: Velocity} shows the maximum velocities attained by the users while achieving targets at different positions. The maximum velocity was found to scale approximately linearly with the target position with average $R^2$ values $53\pm2.1$, $62\pm10.1$, $40\pm14.4$, $63\pm1.6$ for PG, IP, TG, and WR, respectively (averaged $R^2$ of all target widths).Target position significantly ($p<0.001$) affected the maximum velocities. In contrast, target width had no significant effect for all gestures. Table~\ref{tab:p-values} below summarizes the p-values for all the conditions.
  

\section{Discussion}
This paper reports a real-time proportional position control of multiple gestures using a wearable SMG sensor. The proposed wearable SMG system could accurately track various levels of muscle deformation to generate a proportional position control signal. The movement trajectories decoded from the SMG signals exhibited a high degree of congruence with the MCP joint angles and, therefore, had a one-to-one correspondence to the user's volitional motor intent. Our approach quantifies the muscle deformation from sparse SMG signals and maps it to the position of the on-screen user cursor proportionally, as shown in Fig.~\ref{fig: user_trajectories}. Experiment 1 confirmed that there exists a linear relationship between the SMG control signal and the MCP joint angle across all the gestures, as shown in Fig.~\ref{fig: DGVsSMG}. Thus, our results are in agreement with prior B-mode imaging-based SMG studies that reported accurate estimation of joint angles and finger positions by tracking changes in muscle anatomical structures despite having sparse measurement sites~\cite{Akhlaghi7320970}. Therefore, SMG could provide naturalistic control as it measures muscle deformation during dynamic activity.

Functional activities require accurate and stable control over extended periods of time over multiple degrees of freedom (DoFs). So far, wearable SMG systems have been used to demonstrate offline proportional control using muscle contraction levels for up to 8 grasps~\cite{8654210}. Yang et al. also achieved real-time simultaneous proportional position control of the hand and wrist with success rates $>97\%$. However, the stability of the SMG control was evaluated only for a dwell time of 300\,ms. On the other hand, our study systematically evaluates the effectiveness of SMG control for four degrees of freedom. We have not evaluated our system's ability to control the DoFs simultaneously. Fig.~\ref{fig: user_trajectories}shows the trajectories for the entire trial duration of 15 sec. The trajectories show that the participants remained within the target for at least 2\,s upon achievement of the target, resulting in success rates $>80\%$ for higher error tolerance, as seen in Fig.~\ref{fig:successrate}. Additionally, the responsiveness of SMG control can be seen in Fig.~\ref{fig: MovementTime} with an average movement time of $5.6\pm2$\,s. Furthermore, the participants' performance was evaluated by modulating the task's difficulty by changing the target width. The difficulty of the task significantly affected both success rate and movement time. The targets with 5\% width constrained the participants to achieve fine control over the cursor, while the targets with 15\% width allowed the participants to have higher errors between the target and the user cursor. The participants achieved easier targets, i.e., tasks with a target width of 15\%, in significantly lower time than difficult targets (target width of 5\%). However, finer control for lower target widths could be achieved by adapting to the system with sufficient training. 

For proportional positional control, minimal error between the target and end-effector is desirable to achieve intuitive control. The accuracy of the SMG control was evaluated using endpoint error and stability. Fig.~\ref{fig:endpointError} and Fig.~\ref{fig:endpointStability} show that the maximum error between the target and the user cursor and the maximum variance in the user cursor is less than 5\%. Therefore, SMG control seems to minimize error and variability as the participant reaches close to the target. This also suggests that the participants adapted to SMG-based control by trying to achieve the targets with minimum error and variance while learning the non-linearities and inherent noise of the SMG control algorithm, thus retaining their natural motor control abilities~\cite{Todorov2004}. 

\subsection{Limitations and future scope}
There are a few limitations of the presented wearable system. Firstly, the system is not fully wearable in its current state. In the future, a custom-developed wearable pulser-receiver system will be used to make the system fully wearable. In terms of control technique, the proportional control of each gesture was tested in isolation. Control strategies with relevant switching techniques will be developed to achieve sequential or simultaneous control over multiple gestures. 

\section{Conclusion}
A wearable 8-channel SMG sensor array was developed and optimized. A simple technique was developed to derive a real-time proportional positional sonomyography signal from the sensor array. The performance of the optimized sensor array and the derived SMG signal was tested at two locations on the forearm, and the proximal location was chosen due to the high degree of linearity with joint angles across the gestures. Target achievement tasks performance studies performed with healthy individuals demonstrate that the system is able to provide accurate and stable sonomyographic control over multiple degrees-of-freedom. 

\bibliographystyle{IEEEtran}
\bibliography{Bibliography/references}

\end{document}